\documentclass[11pt,aip,apl,preprint,showpacs]{revtex4}
\usepackage{graphicx}
\usepackage{amssymb}
\usepackage[T1]{fontenc}
\usepackage[american]{babel}

\begin{document}

% \title{Fabrication of nanostructured BiFeO3 ferroelectic thin films on Spark plasma sintered LaAlO3 substrates.}
\title{BiFeO$_3$/La$_{0.7}$Sr$_{0.3}$MnO$_3$ heterostructures deposited on Spark Plasma Sintered LaAlO$_3$ Substrates}

\author{D.~Pravarthana$^1$, M.~Trassin$^2$$^,$$^3$, Jiun Haw Chu$^2$}
\author{M.~Lacotte$^1$, A.~David$^1$}
\author{R.~Ramesh$^2$$^,$$^4$$^,$$^5$}
\author{P.A.~Salvador$^6$}
\author{W.~Prellier$^1$}\thanks{wilfrid.prellier@ensicaen.fr} 
\affiliation{$^1$Laboratoire CRISMAT, CNRS UMR 6508, ENSICAEN, Normandie Universit$\acute{e}$, 6 Bd Mar$\acute{e}$chal Juin, F-14050 Caen Cedex 4, France.}
\affiliation{$^2$Department of Physics, University of California, Berkeley, California 94720}
\affiliation{$^3$Department of Materials, ETH, Wolfgang-Pauli-Str. 10, 8093 Zurich, Switzerland}
\affiliation{$^4$Department of Materials Science and Engineering, University of California, Berkeley, California 94720}
\affiliation{$^5$Materials Science Division, Lawrence Berkeley National Laboratory, Berkeley, California 94720}
\affiliation{$^6$Department of Materials Science and Engineering, Carnegie Mellon University, 5000 Forbes Ave., Pittsburgh, Pennsylvania 15213.}

\date{\today}

\begin{abstract}
\quad 

Multiferroic BiFeO$_3$ (BFO) / La$_{0.7}$Sr$_{0.3}$MnO$_3$ heterostructured thin films were grown by pulsed laser deposition on polished spark plasma sintered LaAlO$_3$ (LAO) polycrystalline substrates. Both polycrystalline LAO substrates and BFO films were locally characterized using electron backscattering diffraction (EBSD), which confirmed the high-quality local epitaxial growth on each substrate grain. Piezoforce microscopy was used to image and switch the piezo-domains, and the results are consistent with the relative orientation of the ferroelectric variants with the surface normal. This high-throughput synthesis process opens the routes towards wide survey of electronic properties as a function of crystalline orientation in complex oxide thin film synthesis.

\end{abstract}

\pacs{81.15.Fg, 73.50.Lw, 68.37.Lp, 68.49.Jk}

\maketitle

\newpage

%\section{Introduction}
%%%%%%%%%%%%%%%% INTRODUCTION %%%%%%%%%%%%%%%%%%%%%%%
Transition metal oxides display many exciting physical phenomena, including ferroelectricity in perovskites, high-T$_C$ superconductivity in layered cuprates, colossal magnetoresistance (CMR) in perovskite manganites, as well as the coexistence of magnetism and ferroelectricity (i.e., multiferroicity) observed in iron-based materials.\cite{cheong} Interest in thin film transition metal oxides is driven in part by the potential technological application of devices exploiting such phenomena, i.e. oxide electronics, and in part by the novel structures and properties observed in epitaxial oxide films, using phase, strain, and interfacial engineering.\cite{metastable,film}
In spite of the large number of observations and promise of epitaxial oxide thin films, prior investigations have largely focused on films on low-index commercially-available single-crystal substrates. This limits the investigations of growth, which is dependent on substrate surface, and understanding of anisotropic functional properties (i.e., polarization, magnetization, resistivity,  etc.),\cite{anisotropy2,anisotropy3,Shekhar} which are dependent on the orientation of the film with respect to the surface, to a narrow region of special interfaces/orientations. Even though, general high-index $(hkl)$ single crystals can be used for growth \cite{621}, their cost and availability limit detailed investigations. To investigate the structure-property relationships in thin films over all the orientation space requires a combined and systematic high-resolution analysis of both features at the grain scale in samples.

Here, we develop further a high-throughput synthesis process (called combinatorial substrate epitaxy, or CSE) where an oxide film is grown epitaxially on a polycrystalline substrate, \cite{Phase, CSE1, CSE2, CSE3, CSE4} which should allow functional properties to be investigated across the entirety of epitaxial orientation space. Prior work using CSE has focused on understanding growth of non-isostructural films substrate pairs \cite{Phase, CSE1, CSE3, CSE4}, engineering phase stability in complex materials,\cite{CSE2} and establishing the role of substrate and film phase/orientation on photochemistry.\cite{photochem} Complex oxide multilayer (heterostructures) films exhibiting functional properties have not been investigated using CSE. In CSE, each grain of the polycrystalline substrate can be viewed as a single crystal substrate with a specific crystallographic orientation, and there are thousands of substrates in any given film deposition. The orientation and structural quality of the substrate and film grains are investigated locally (and mapped) using Electron Backscattering Diffraction (EBSD). The physical property of the films can also be mapped using local probes, such as scanning probe and near-field microscopy methods, but has not been carried out to date for functional electronic properties. If CSE can be coupled with scanning property measurements, high-throughput correlations can be generated between functional properties, film-substrate pairs, and crystal orientation, providing a library of physical property observations and expanding our understanding of engineering function into transition metal oxides.

To illustrate this combined approach, heterostructures of conducting La$_{0.7}$Sr$_{0.3}$MnO$_3$ (LSMO) and multiferroic BiFeO$_3$ (BFO) films were grown, using pulsed laser deposition, on polished polycrystalline LaAlO$_3$ (LAO) ceramic substrates, fabricated using spark plasma sintering (SPS). Using similar heterostructures, the ferroelectric and magnetic properties of BiFeO$_3$ thin films have been widely studied on low-index orientations, such as (100), (110), and (111), of single crystal perovskites, such as LaAlO$_3$, NdGaO$_3$, SrTiO$_3$, and DyScO$_3$, showing that the properties of the films were dependent on both the orientation of the substrate and the strains.\cite{BFO,BFO1,BFO2,manoj} 
Owing to the orientation dependent properties observed on low-index single crystals, it is of interest to determine whether the same effects are observed in BiFeO$_3$ grains deposited on polycrystals having similar low-index orientations, as well as whether the functional properties are observed locally everywhere in a locally epitaxial polycrystalline film. The epitaxial relationships of BiFeO$_3$ films on LaAlO$_3$ are obtained using EBSD, while the local ferroelectric properties of each grain are probed using piezoforce microscopy (PFM). We demonstrate that the local properties of BiFeO$_3$ prepared using CSE are consistent with those expected for typical single crystals, and these combined methods open a door to synthesize and screen various electronic properties as a function of crystallographic direction using a single sample.

Equimolar amounts of undoped commercial $\alpha$-Al$_2$O$_3$ and La$_2$O$_3$ powders (La$_2$O$_3$, Aldrich and Al$_2$O$_3$, Cerac with 99.9 \% purity) were weighted in stoichiometric proportions, mixed intimately and reacted in their solid states using different thermal treatments. The precursors were finally calcined 12 hours at 1200 $^{\circ}$C to obtain the desired phase.\cite{pravat} Another step of grinding was necessary to obtain powders with homogeneous grain sizes. The resulting LaAlO$_3$ powders were sintered under 50 MPa at 1550$^{\circ}$C for 30 minutes using Spark Plasma Sintering (SPS), as described elsewhere.\cite{pravat} The ceramics were confirmed to be LaAlO$_3$ using conventional x-ray diffraction. Substrates were cut from the SPS ceramic, were mechanically polished to 10 $\mu m$ using silicon carbide paper, and were successively polished with diamond pastes of 3 $\mu m$ and 1 $\mu m$ for about 2 minutes, resulting in a mirror-like surface with a roughness of 3-4 nm.\cite{pravatox} The polished surface was finally etched in 5 \% HF:HNO$_3$ solution to remove surface contaminants and to release strain due to polishing. 
The heterostructures consisted of a 100 \AA~ thick LSMO base layer and a 500 to 1000 \AA~ thick BFO top layer, where the LSMO served as an epitaxial metallic bottom contact for the piezoresponse force microscopy (PFM) analysis.\cite{LSMO} Both film layers were grown by pulsed laser deposition, using a KrF excimer laser ($\lambda$=248 nm) with a fluence of 1 J/cm$^2$ and a repetition rate of 2 Hz. For all depositions, the substrate temperature was kept at 700 $^{\circ}$C under a 100 mTorr O$_2$ partial pressure. 

Structural and microstructural characterization of the ceramics and films were carried out using electron backscatter diffraction (EBSD). The samples were mounted at a 70$^{\circ}$-tilt angle from horizontal in a scanning electron microscope (SEM) operated at 20 kV. The epitaxial relationships between film and substrate were also obtained.\cite{Phase,CSE3} Kikuchi patterns were recorded and indexed automatically by the EDAX orientation imaging microscopy (OIM$^{TM}$) software (v.6), which has a typical resolution of 2-10 nm in-depth and 30-90 nm in lateral (meaning the BiFeO$_3$ film are thick enough to have no extraneous contributions to the patterns from the under layers).\cite{kikuchi depth} Both patterns were indexed using the Pm-3m space group with lattice parameters of 3.8 \AA~(3.96 \AA) for the LAO substrate (BFO film) for clarity and for direct comparison with literature.\cite{LAO,BFO} 

The confidence Index (CI = 0-1), which measures the difference between the best and second best indexing solutions based on a voting algorithm is used to describe the EBSD data. To get more reliable data, points with poor or erroneous indexing (< 0.1) have been removed, in particular near grain boundaries. By scanning the surface of the sample with a beam step size of 0.3 $\mu$m, several Kikuchi diagrams are recorded. The software assigns a color pixel for each orientation, and inverse pole figure (IPF) maps of the surface of the substrate and the film are thus recorded. Piezoelectric properties (PFM analysis) were characterized by an AFM-based setup (Veeco-DI, equipped with a Nanoscope IV controller) under ambient conditions using a commercial TiPt-coated silicon tip.

Orientation maps plotted as inverse pole figures (IPFs) are shown in Fig.1a and 1b for a LAO substrate and a BFO film deposited on this substrate (with an LSMO buffer layer), respectively. The color code for the IPFs is given Fig.1c. The images in Fig.1a and 1b were recorded from approximately the same areas, and the white squares highlight identical grain areas in the two images. %PAS 1. The color code has some residual uncropped dark spots below it, should be cleaned up. 2. Why does the image in b look less pixelated? 3. you should put the letters (a) and (b) inside the plot, the lost information is not that critical here. That way they will be larger in the actual publication. 4. There should be a scale bar in image (b), or a comment that they are the same. They seems slightly different, but that could be a pixelation issue. 5. It is unfortunate that the white box has a grain with the green cross on it (i assume this is a marker) that does not show up as the same color in the film... It distracts from the point that most grains do have the same color, but I see why the box is where it is...
 The LAO substrate has a bimodal grain size distribution, with some very large grains > 150 $\mu m$ and smaller grains ranging from 50-100 $\mu m$.  While all colors can be observed in the IPF map, there is a large area fraction of purple and blue grains, indicating the grains have non-random texture in the substrate; the detailed grain and texture analysis of the SPS prepared LAO will be published elsewhere.\cite{pravatox} %PAS This image does not support a 111 texture, purple is by far the largest area fraction... which is not 111. It is not worth splitting hairs about whether it is 111 or 211 at this point, so it is not worth distracting the reader.
 The film grains have very consistent assignments for the orientations within almost all grains of the substrate, such that the cleaned data exhibits grain sizes and grain boundary locations that are very similar in the two images, indicating that both LSMO and BFO films grew in a grain-over-grain fashion on the LAO substrate. The average orientation (as assigned by the software) of each film grain is nearly identical to the orientation of the substrate grain on which it grew. Color variations between the film and substrate (see the purple/pink pair within the white boxes) are usually within a few degrees of each other on in the color key shown in Fig.1c, and small angular misorientations such as these have been observed previously in CSE grown films\cite{CSE2, CSE4} and are expected to relieve misfit strains in the heterostructures.\cite{misorient} The in-plane alignments between the substrate and grain exhibited identical characteristics to the out-of-plane alignments discussed here. These observations indicate that the film epitaxy is dominated by local substrate-driven growth events that are consistent with each perovskite layer adopting a cube-on-cube epitaxial orientation relationship, for nearly all grain orientations of the substrate.
  
EBSD patterns were recorded for each grain (but not shown). The bands of the LaAlO$_3$ substrate were sharp and intense, while the BiFeO$_3$ film values are slightly more diffuse than the substrate, which can be attributed to local strains in the epitaxial films or the inhomogeneous strains in the film resulting from relaxation phenomena.\cite{CSE1,CSE2,film,strain1} 
 Nevertheless, all grains investigated display otherwise similar patterns with good CI values, attesting a good crystalline quality, and confirming the epitaxial perovskite phase formation for the film. The local epitaxial quality was further confirmed by comparing the misorientation angle between the individual local orientations in a given grain to the average orientation of that grain (the average values were plotted in Fig.1d). A plot of this misorientation angle is shown in Fig.1d for the BFO film, taken from the region highlighted in Fig.1b by the white box.  
 For most of the grains in the image (the grain near (111) and (110) are marked), the misorientation angle is within 1$^{\circ}$, in agreement with expected values from epitaxial perovskite heterostructures. A similar analysis performed over many grains in the film indicates the majority of grains are of such quality, except for the (001) oriented film grains.
Film growth on LAO grains near (001) tends to result in the formation of small clusters that exhibit a local misorientation angle ranging from 3-5$^{\circ}$ from the average (Fig.1d). The IQ map shown in Fig.1e further highlights the multiple orientations within (001) oriented grains. The IQ parameter is the integrated intensity over all peaks in the Hough Transform, which reflects the quality of the local diffraction pattern and can be correlated to the phase, orientation, and strain in the diffraction volume.\cite{Image quality3} IQ maps have high contrast between grains and where local orientation/strain variations exist, such as at grain boundaries. In Fig.1e (dark (light) regions are associated to low (high) image qualities), the IQ contrast varies in strong correlation to the regions of local misorientation observed in Fig. 1d. Fig.1d and 1e show that the (001) grains are non-homogeneous in colour, while (111) and (101) are homogeneous, as are the films on higher index orientations. It is intriguing how uniform the vast majority of the BFO grains are in these heterostructures, and also that the BFO films on (001) grains have multiple orientations that vary in absolute orientation by about 5$^{\circ}$, when indexed in the cubic system. 
%PAS I am not sure I am thinking about strain in the same fashion as you are. I am always worried about mixing up homogeneous (coherency) strains and inhomgeneous (dislocation) strains in cubic materials, but also twinning and phase related strains in non-cubic (especially LAO) systems. It is true that the IQ map has strain information, but there is not one flavor of strain... For example, a perfectly coherent but strained films should have a high and uniform image quality. A heavily dislocated (relaxed) but uniform film should have a low and uniform image quality. A coherently strained, but twinned, film might have an intermediate and non uniform IQ, and a relaxed and twinned (or multiphase) would have a low and non-uniform IQ. The multiphase nature of the (001) oriented film T and R should be enough to indicate that the IQ would vary greatly. The fact that there are low angle boundaries (3-5 �) is also enough to cause the IQ map to look this way. So, i took out the ambiguous references to strain. 
 
%%%%%%%%%%%%%%%% FIGURE 2 %%%%%%%%%%%%%%%%%%%%%

These observations indicate that one can investigate film growth on different orientations by exploring specific grains in a polycrystalline matrix, similar to what has been done extensively using single crystals.\cite{BFO} To explore this further, Kikuchi patterns were recorded from the two different kinds of regions observed in Fig.1d and e for the (001)-oriented grains.  The Kikuchi pattern registered from the low misorientation (high IQ) region is shown in Fig.2a. The Kikuchi pattern registered from the high misorientation (low IQ) region is shown in  Fig. 2b and it clearly differs from that shown in Fig.2a. We propose that this region is associated to the tetragonal (T)-like phase observed in BFO films on (001) pseudo-cubic substrates,\cite{Mixed phase} and this would explain the strong variation in assigned orientation and image quality observed in Fig.1. Similar patterns were always obtained for these two types of regions on grains near the (001), indicating the growth is uniform and that the mixed phase growth is a common feature in films, but only near (001). %PAS I think the argument in a bit weak, but I like it. Can you make the argument stronger by explaining the Kikuchi patterns better and why they (1) look like they do and (2) why they correspond to the phase they do. Can the tetragonal phase be relaxed or is it always coherent? How did the poles get labeled, they were not discussed and they are very different. 

%%%%%%%%%%%%%%%% FIGURE 3 %%%%%%%%%%%%%%%%%%%%%

The epitaxial BFO heterostructure was investigated using PFM, in a similar fashion to BFO films grown on single crystals.\cite{BFO,BFO1,BFO2} The grains highlighted in the white box in Fig.1 were first investigated. Fig. 3a and 3b show the crystalline orientation from EBSD, where specific orientations and misorientation angles are marked, and the corresponding topography obtained by PFM, respectively. The topography is observed to change from one grain to another, and the root-mean-square (RMS) roughness values range from 1 to 10 nm, depending on the grain. To further analyze the influence of the substrate grain orientation on the multiferroic thin film properties, the ferroelectric domains were imaged using PFM. Fig.3c and 3d show the out-of-plane (OP) and in-plane (IP) component of the PFM response, respectively. As the variation in PFM contrast shows,  the ferroelectric architecture is clearly correlated with the underlying grain structure of the substrate, and changes dramatically exactly at the grain boundaries. The BiFeO$_3$ compound has been widely studied, it now well established that (111) oriented film can have only two possible polarization direction, i.e. upwards or downwards without any in-plane component.\cite{M1} For the considered grain, the out-of-plane component of the polarization is pointing upwards on the (111) oriented grains (darker contrast in Fig.3c) and downwards on the (001) oriented grain (white OP contrast in Fig. 3c). The changes in the OP polarization component can be attributed to the different LSMO growth modes on the substrates, leading to different electrostatic environments for the ferroelectric layer,\cite{morgan1,morgan2} and is shown in supplementary materials (figure S1). Correspondingly, the in-plane (Fig.3d) polarization component reflects the underlying grain distribution and shows that all 4 possible BFO ferroelectric polarization variants are present (with dark, bright and no contrast for up, down and left/right polarization direction, respectively. We note that this large scanning area, showing different grains, allows for a qualitative analysis and cannot evidence nanostructured domains. 

To further probe the properties of the film on the SPS prepared LAO substrates, we investigated whether the ferroelectric domains could be switched. Fig.4a-c shows the topography, out-of-plane PFM and in-plane PFM images, respectively, after the local application of a positive 12 V bias in the central 20x20 $\mu m^2$ region that includes several grains and grain boundaries. A clear polarization reversal (comparing the central region to the outer regions) is observed in the OP image (Fig.4b), as the PFM contrast is reversed. The observed change in the out of plane contrast with a specific voltage polarity (inducing up to downward OOP switching only in our configuration) further validate the interpretation of the direction of the OOP polarization component. This demonstrates that the ferroelectric properties of BFO are maintained for these many orientations in the heterostructure on the LAO substrate. The in-plane PFM contrast evinces the different local switching characteristics, as the contrast only reverses in some BFO films grown on specific grains. These different switching behaviors are due to the correlation between the allowed ferroelectric variants and the local grain orientation.\cite{M1}To demonstrate further the high-quality local ferroelectric properties, a (110)-oriented grain was selected and the BFO's ferroelectric polarization was switched using the PFM. The topography of this grain is shown in Fig.4d, and the typical stripes expected from a (110)-oriented film are observed.\cite{M1,New} In this configuration, only 180 degree switching events are possible using our tip bias value.\cite{M2} An horizontal 500 nm wide line was written in this grain using a + 12 V bias, and the OP and IP PFM images are shown in Fig.4e and 4f, respectively. 180 degree switching was observed as both the OP and IP contrast reversed (the underlying stripe structure is attributed to the topography contribution), again demonstrating that the polycrystalline substrate grains are similar to microcrystalline single crystal surfaces, as the properties of the multiferroic heterostructures are similar to those expected from single crystals.

In summary, it was demonstrated that structure-property relationships can be investigated for complex oxide ferroelectric heterostructures using the combinatorial substrate epitaxy approach, a high-throughput method for investigating the local epitaxy and properties of films deposited on polycrystalline substrates. High quality BiFeO$_3$ and La$_{0.7}$Sr$_{0.3}$MnO$_3$ thin film heterostructures were deposited by pulsed laser deposition on dense LaAlO$_3$ ceramics prepared by spark plasma sintering. The structural quality of the substrate and BFO films, as well as the epitaxial relationships of the films to the substrate, were determined locally using electron backscatter diffraction. For all but the (001) orientated substrate grains, the all perovskite heterostructure exhibited (so-called) cube-on-cube orientation relationships with misorientations between layers of less than 1$^{\circ}$. On (001) oriented LaAlO$_3$ grains, the misorientation values were between 3-5$^{\circ}$, and were attributed to the presence of two BFO phases. The presence and switchability of piezo-domains was evinced using PFM, confirming that BFO films are ferroelectric when locally epitaxial on polycrystalline substrates. Ultimately, we obtain a clear correlation between the grain orientation and the intensity of the piezo-domains, all of which are consistent with relative orientation of the ferroelectric variants with the surface normal. The wide variety of crystal orientations available in the polycrystalline substrate opens a high throughput route for establishing libraries of specific properties as a function of orientation in oxide films. Though well-known all-perovskite heterostructures were used in this study, the investigation of the epitaxial growth and structure-property relationships are now possible for multifunctional oxides with complex atomic structures, such as GaFeO$_3$ or Gd$_2$Mn$_2$O$_5$, for which commercially available isostructural substrates are not available.

We thank L. Gouleuf and J. Lecourt for technical support. D.P. is supported by a PhD fellowship included in the Erasmus Mundus Project IDS-FunMat. M. Lacotte received her PhD scholarship from the Minist$\grave{e}$re de l'Enseignement Sup$\acute{e}$rieur et de la Recherche. Partial support of the French Agence Nationale de la Recherche (ANR), through the program Investissements d'Avenir (ANR-10-LABX-09-01), LabEx EMC3 and the Interreg IVA MEET project is also acknowledged. We also thank O. Copie, R. de Kloe, I. Canero Infante, J. Wang and R. Ranjith for fruitful discussions.

\newpage
Figures Captions

%Figure 1: Typical Kikuchi patterns recorded for (a) LaAlO$_3$ substrate, and (b) BiFeO$_3$ film. The orientation with respect to the surface plane determined from the measured Euler angles in standard angle representation ($\varphi_1, \phi, \varphi_2$) for substrate (18,75,51) and film (146,34,284)

Figure 1: Inverse pole figure (orientation) maps of (a) the LaAlO$_3$ substrate and (b) the BiFeO$_3$ film, and their corresponding colour code (c). The white boxes indicate grains in the same region, as well as the region from which the maps in (d) and (e) were produced. (d) A the map of the local misorientation angle relative to the average surface orientation of each grain and (e) a corresponding grey scale image quality (IQ) map from the same region. In (d), blue corresponds to minimum value of 1 and 5 for maximum in red colour.

%PAS I am assuming this misorientation angle is relative to the average of the grain, not to the substrate.

Figure 2: Two types of kikuchi patterns seen on (001) oriented grains of BiFeO$_3$ film with (a) : Rhombohedral symmetry and (b) : Tetragonal symmetry. The orientation with respect to the surface plane determined from the measured Euler angles in standard angle representation ($\varphi_1, \phi, \varphi_2$) for High IQ and Low IQ regions are (40, 74, 285) and (89, 28, 315), respectively.

Figure 3: Enlarged mapping on BiFeO$_3$ (001)-oriented grain. (a) : EBSD colour code orientation map, (b) : PFM Topography, (c) : Out-of-plane PFM and (d) : In-plane PFM.

Figure 4: Polling measurements of BiFeO$_3$ films across triple junctions (a-c) and grain boundaries (d-f). (a) : PFM Topography, (b) : Out-of-plane ,(c) : In-plane, (d) : PFM Topography, (e) : Out-of-plane and (f) : In-plane. PFM contrasts were recorded after local application of a +12 V bias showing a 180 degree switching event. 
%Figure 4: PMF measurements of BiFeO$_3*$ films. (a) EBSD map   of substrate with size ranging from 0.5-10 micron grain size, (b) PFM Topography, (c)  Out-of-plane ,(d) In-plane, and (e) SEM microstructure nm with bottom electrode LSMO exhibit formation of grooves due to in plane strain.

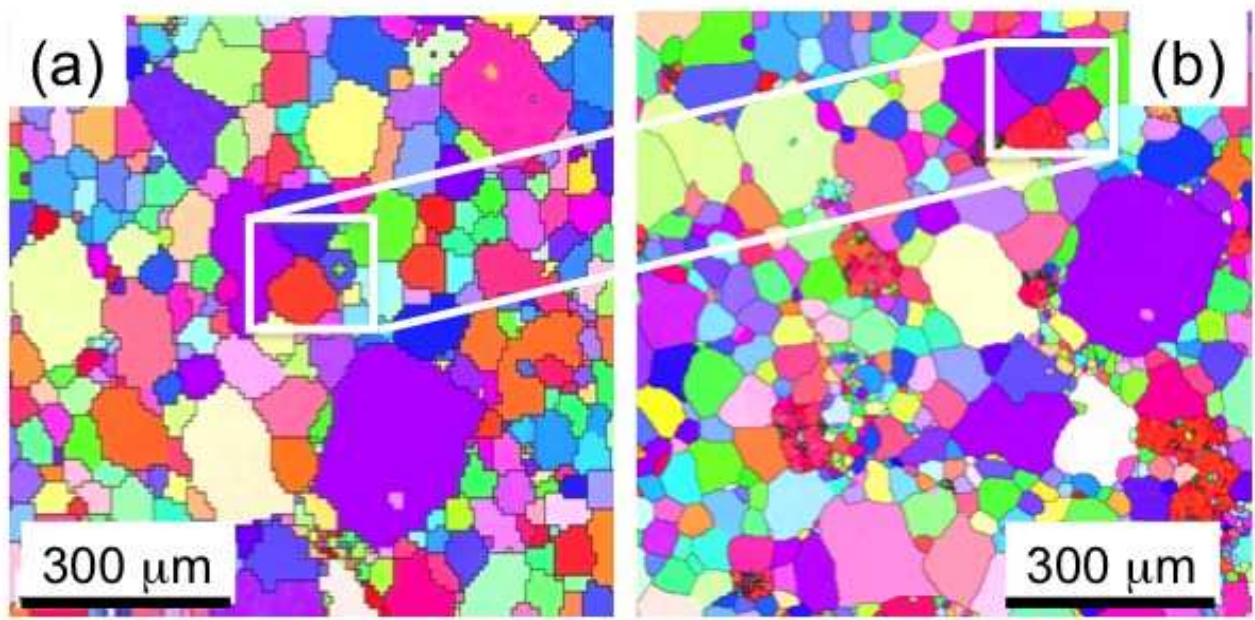
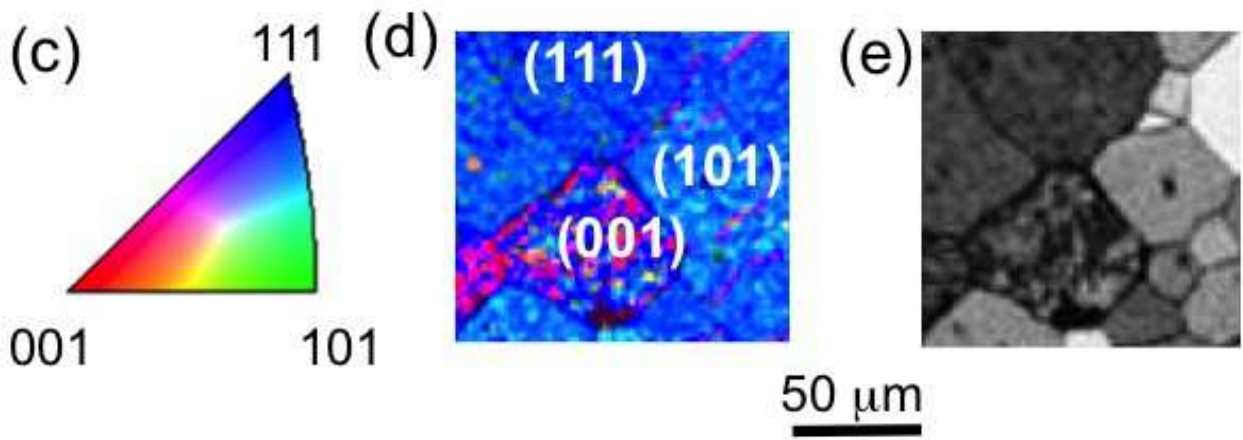
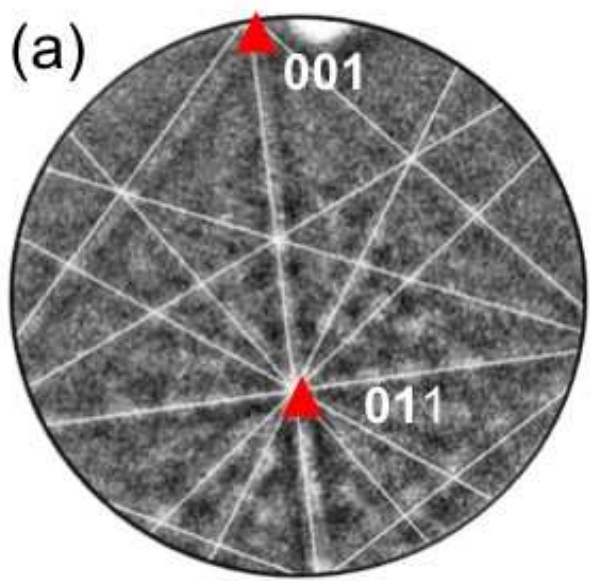
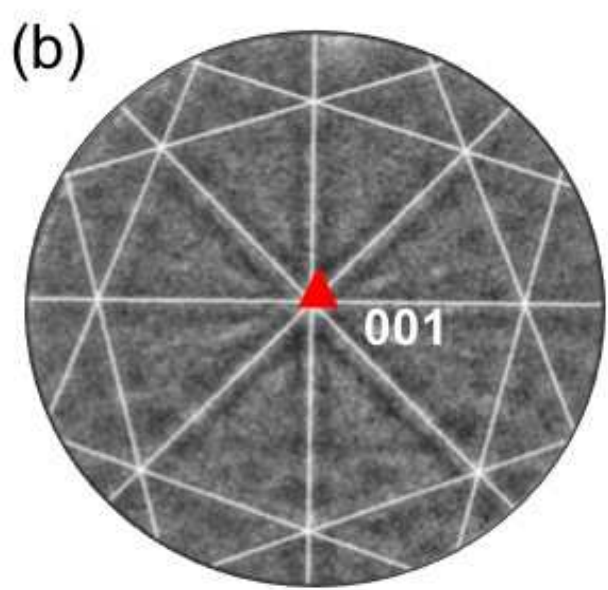
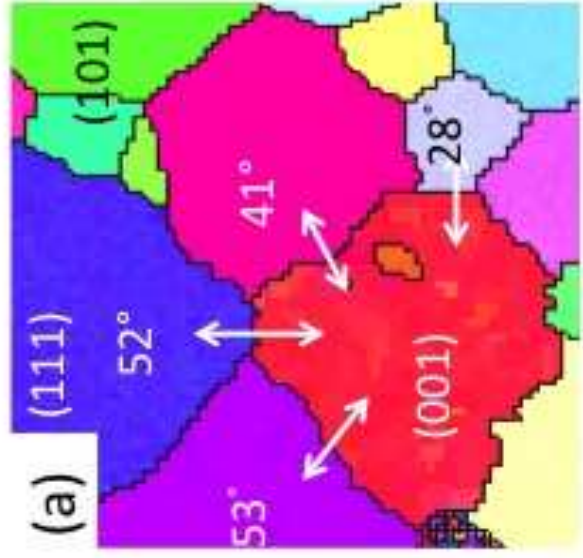
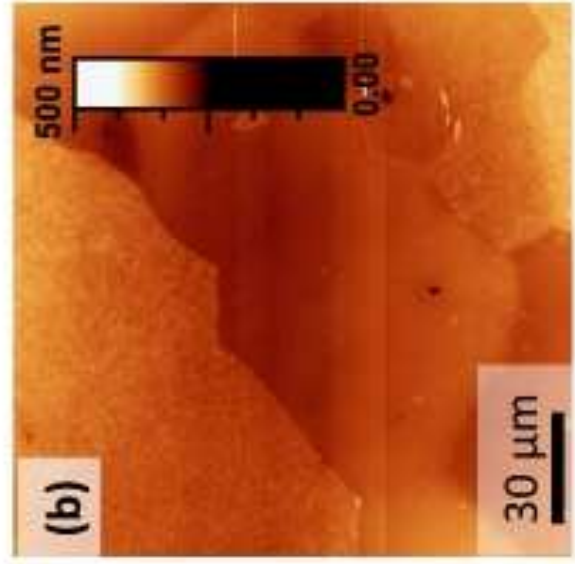
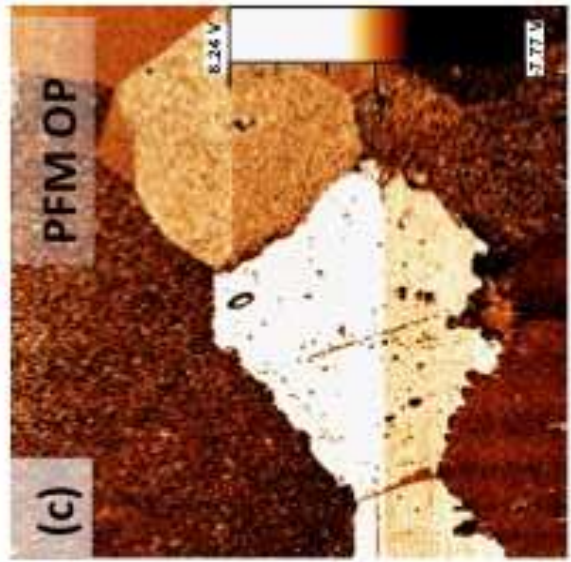
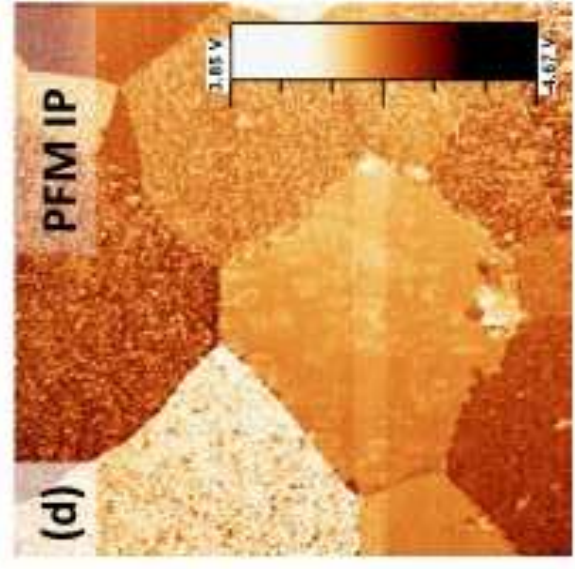
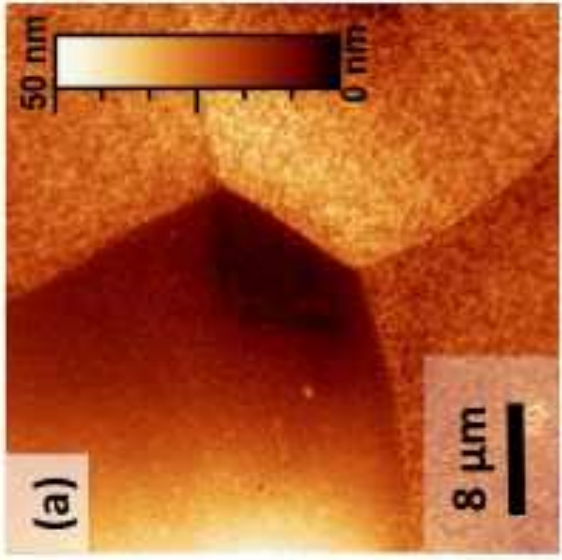
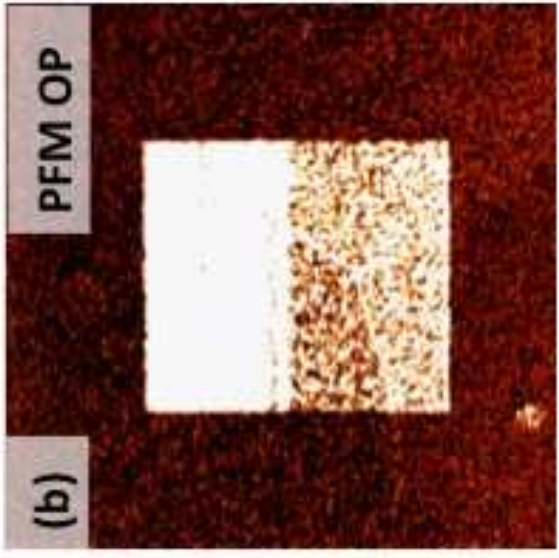
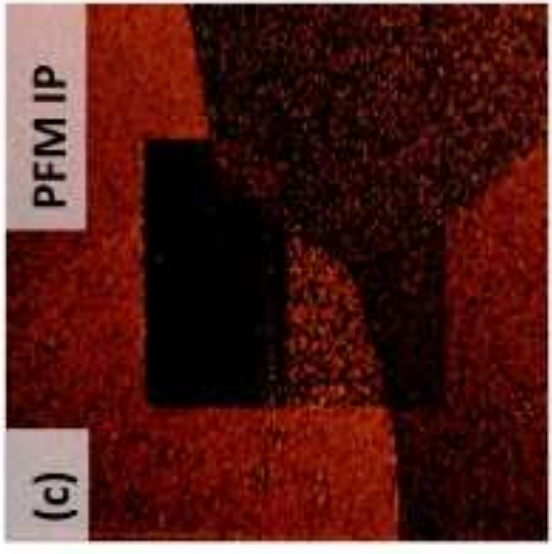
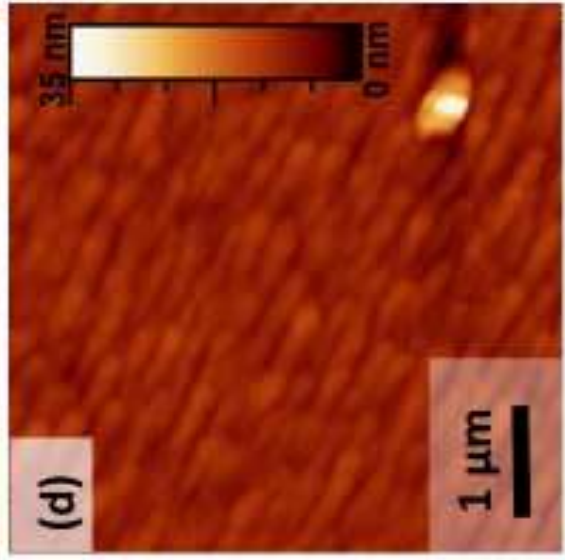
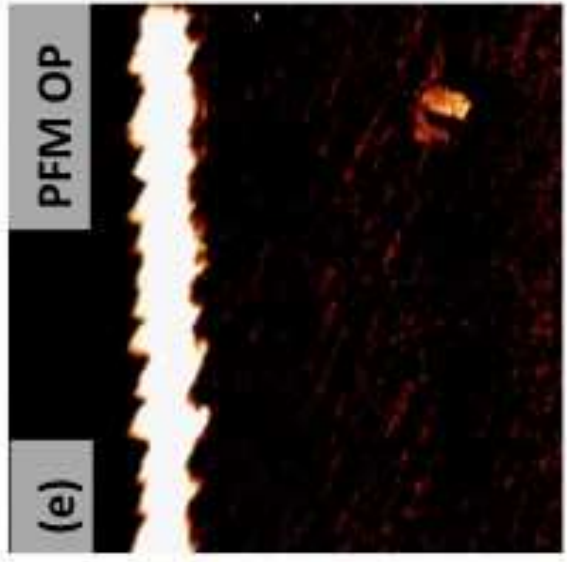
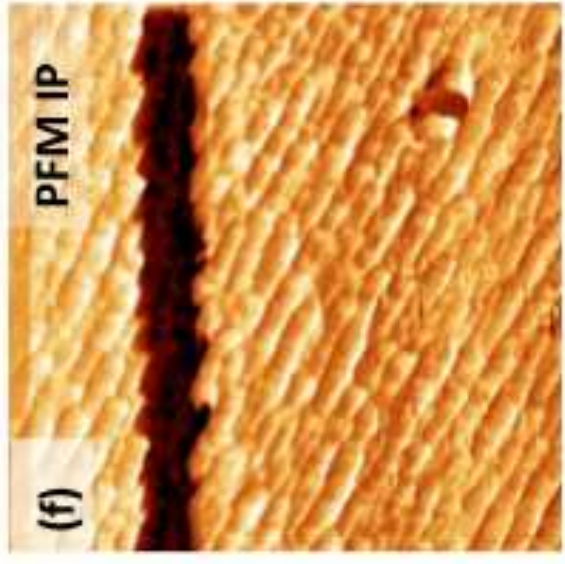
\end{document}